\documentclass[prd,nofootinbib,superscriptaddress,twocolumn]{revtex4}
\usepackage{amsfonts,amssymb,amsthm,bbm,amsmath}
\usepackage{hyperref}

\usepackage{color}

\usepackage{tikz}
\usetikzlibrary{calc}
\usetikzlibrary{decorations.pathmorphing}
\usepackage{pict2e}
\makeatletter
\DeclareRobustCommand{\loplus}{\mathbin{\mathpalette\dog@lsemi{+}}}
\DeclareRobustCommand{\lotimes}{\mathbin{\mathpalette\dog@lsemi{\times}}}
\DeclareRobustCommand{\roplus}{\mathbin{\mathpalette\dog@rsemi{+}}}
\DeclareRobustCommand{\rotimes}{\mathbin{\mathpalette\dog@rsemi{\times}}}

\newcommand{\dog@rsemi}[2]{\dog@semi{#1}{#2}{-90,90}}
\newcommand{\dog@lsemi}[2]{\dog@semi{#1}{#2}{270,90}}
\newcommand{\dog@semi}[3]{%
  \begingroup
  \sbox\z@{$\m@th#1#2$}%
  \setlength{\unitlength}{\dimexpr\ht\z@+\dp\z@\relax}%
  \makebox[\wd\z@]{\raisebox{-\dp\z@}{%
    \begin{picture}(1,1)
    \linethickness{\variable@rule{#1}}
    \roundcap
    \put(0.5,0.5){\makebox(0,0){\raisebox{\dp\z@}{$\m@th#1#2$}}}
    \put(0.5,0.5){\arc[#3]{0.5}}
    \end{picture}%
  }}%
  \endgroup
}
\newcommand{\variable@rule}[1]{%
  \fontdimen8  
  \ifx#1\displaystyle\textfont3\else
    \ifx#1\textstyle\textfont3\else
      \ifx#1\scriptstyle\scriptfont3\else
        \scriptscriptfont3\relax
  \fi\fi\fi
}
\makeatother
\definecolor{airforceblue}{rgb}{0.36, 0.54, 0.66}
\definecolor{antiquefuchsia}{rgb}{0.57, 0.36, 0.51}
\definecolor{blush}{rgb}{0.87, 0.36, 0.51}
\definecolor{bondiblue}{rgb}{0.0, 0.58, 0.71}
\definecolor{MyGreen}{rgb}{0.0,0.5,0}
\definecolor{MyDarkRed}{rgb}{0.7,0,0}
\definecolor{MyBlue}{rgb}{0.0,0.0,.8}
\definecolor{Green}{rgb}{0.4,.8,0}
\def\be#1\ee{\begin{align}#1\end{align}}
\def\bsub#1\esub{\begin{subequations}#1\end{subequations}}
\def\bg#1\eg{\begin{gather}#1\end{gather}}
\def\ba{\begin{eqnarray}}
\def\ea{\end{eqnarray}}

\def\nn{\nonumber}
\def\q{\qquad}
\def\f{\frac}
\def\df{\dfrac}

\def\rm#1{\mathrm{#1}}
\def\lp{\ell_\text{Pl}}
\def\ls{\ell_\text{s}}

\def\de{\mathrm{d}}
\def\rd{\mathrm{d}}

\def\pp{\partial}
\def\tV{\tilde{V}}
\def\ttau{\tilde{\tau}}

\newcommand{\C}{{\mathbb C}}

\newcommand{\R}{{\mathbb R}}
\newcommand{\Z}{{\mathbb Z}}

\newcommand{\kC}{{\mathfrak C}}
\newcommand{\cc}{{\mathfrak c}}

\newcommand{\cH}{{\mathcal H}}
\newcommand{\Hr}{{H_{\mathrm{r}}}}

\newcommand{\cK}{{\mathcal K}}
\newcommand{\cL}{{\mathcal L}}

\newcommand{\cO}{{\mathcal O}}

\newcommand{\cS}{{\mathcal S}}
\newcommand{\cT}{{\mathcal T}}

\newcommand{\cV}{{\mathcal V}}

\newcommand{\Spin}{\mathrm{Spin}}
\newcommand{\SL}{\mathrm{SL}}
\newcommand{\SO}{\mathrm{SO}}

\newcommand{\ISO}{\mathrm{ISO}}

\renewcommand{\sl}{{\mathfrak{sl}}}
\newcommand{\so}{{\mathfrak{so}}}

\newcommand{\iso}{{\mathfrak{iso}}}

\newcommand{\h}{{\mathfrak h}}

\newcommand{\Vone}{{V_1}}
\newcommand{\Vtwo}{{V_2}}
\newcommand{\Vi}{{V_i}}
\newcommand{\Pone}{{P_1}}
\newcommand{\Ptwo}{{P_2}}
\newcommand{\defo}{{\lambda}}

\def\rd{\textrm{d}}
\begin{document}

\title{Regularized Black Holes from Doubled FLRW Cosmologies}
%\title{Deformation of black hole minisuperspace algebra}

\author{{\bf Marc Geiller}}
\email{marc.geiller@ens-lyon.fr}
\affiliation{ENS de Lyon, Laboratoire de Physique, CNRS UMR 5672, Lyon 69007, France}
\author{{\bf Etera R. Livine}}
\email{etera.livine@ens-lyon.fr}
\affiliation{ENS de Lyon, Laboratoire de Physique, CNRS UMR 5672, Lyon 69007, France}
\author{{\bf Francesco Sartini}}
\email{francesco.sartini95@gmail.com}
\affiliation{Esprit des Lieux, Jardin Singulier, Saint L\'eger du Ventoux 84390, France}

\date{\today} 
 
\begin{abstract}

Reduced general relativity for four-dimensional spherically-symmetric stationary space-times, more simply called the black hole mini-superspace, was shown in previous work to admit a symmetry under the three-dimensional Poincar\'e group
%$\Spin(2,1)\ltimes \R^{3}$
$\ISO(2,1)$. Such a non-semi-simple symmetry group usually signals that the system is a special case of a more general model admitting a semi-simple Lie group symmetry. We explore here possible modifications of the Hamiltonian constraint of the mini-superspace. We identify in particular a continuous deformation of the dynamics that lifts the degeneracy of the Poincar\'e group and leads to a $\SO(3,1)$ or $\SO(2,2)$ symmetry. This deformation is not related to the cosmological constant. We show that the deformed dynamics can be represented as the superposition of two non-interacting homogeneous FRW cosmologies, with flat slices filled with perfect fluid. The resulting modified black hole metrics are found to be non-singular.

\end{abstract}

\maketitle  

\makeatletter

\makeatother

%%%%%%%%%%%%%%%%%%%%%%%%%%%%

%%%%%%%%%%%%%%%%
%\subsection*{Introduction}
%%%%%%%%%%%%%%%%

Quantum physics and group theories are strongly connected. The representation theory of symmetry groups gives a powerful handle on quantization. On the road towards a full comprehension of quantum gravity, it is crucial to understand the role of symmetries in general relativity.

There has recently been increasing hints of non-trivial symmetries of general relativity, beyond its gauge invariance under space-time diffeomorphisms, for instance by looking at boundary conserved charges and asymptotic symmetry group, e.g. \cite{Strominger:2021mtt,Freidel:2021fxf,Freidel:2021qpz,Freidel:2021ytz}, or dynamical symmetries of black holes' quasi-normal modes, e.g. \cite{Hui:2021vcv,BenAchour:2022uqo,Hui:2022vbh,Berens:2022ebl}.
In this context, recent works have uncovered the existence of hidden symmetries for cosmological and black hole mini-superspaces, that is the reduction of general relativity to homogeneous space-time or spherically-symmetric metrics. These reduced gravitational systems can be written as mechanical models with finite number of degrees of freedom. Systematically investigating their conserved charges, it was found that these mini-superspaces exhibit symmetries beyond the expected residual diffeomorphism invariance or metric isometries  \cite{BenAchour:2019ywl, Geiller:2020xze,  Sartini:2022ecp, BenAchour:2022fif, BenAchour:2023dgj,Geiller:2022baq}.

In this short paper, we  focus on the case of spherically-symmetric metrics, for which the existence of a group of symmetries isomorphic to the three-dimensional Poincar\'e group $\ISO(2,1)$ has been recently uncovered \cite{Geiller:2020xze, Sartini:2022ecp, BenAchour:2022fif,BenAchour:2023dgj}. The Noether charges induced by this symmetry allow to integrate the dynamics of the system and lead back to the Schwarzschild metrics with arbitrary mass, as expected. 
The presence of a non-semi-simple algebra symmetry suggests the existence of a hidden parameter that givesnback the Poincar\'e algebra in a particular limit, in analogy to what happens when we set the cosmological constant to zero in the study of asymptotic symmetries of general relativity.
%what happens when we set the cosmological constant to zero in the study of asymptotic symmetries in three-dimensional gravity. There, the double copy of the Virasoro group degenerates into the non-semisimple BMS group \cite{Barnich:2012aw}.

In this paper, we proceed to  a systematic investigation of possible deformations of the Hamiltonian constraint of the black hole mini-superspace, and we show that it is indeed possible to identify a deformation parameter $\defo$ such that the symmetry group of the mini-superspace is ``regularized'' to $\SO(3,1)$ for $\defo>0$ and $\SO(2,2)$ for $\defo<0$, while leading back to the Poincar\'e group $\ISO(2,1)$ in the degenerate limit $\defo\to0$. It is important to stress that this deformation parameter $\lambda$ has no link whatsoever with the cosmological constant. Indeed, it does not parametrize a deformation of the space-time metric but of the metric in field space.

%deform its dynamics in a way that the integration of the dynamics corresponds to the flow of a semisimple group. The sign of the deformation parameter $\defo$ selects respectively a double copy of the conformal $\SL(2,\R)$ group for negative values of $\defo$. Conversely, if $\defo>0$ the hidden symmetry becomes the four-dimensional Lorentz group $\SO(3,1)$. The case $\defo=0$ is finally degenerate and corresponds to the original group $\ISO(2,1)$.

The negative deformation parameter case is of particular interest. The factorization of $\SO(2,2)$ as the direct product of two copies of $\SO(2,1)$ allows to map the black hole mini-superspace as the superposition of two independent FRW cosmologies. This surprisingly leads to singularity-free modified black hole solutions. This might suggest a more general basis for singularity resolution in general relativity through symmetry considerations.

We conclude the paper with a discussion of the implications  of the whole family of deformations of the black hole mini-superspace and outlook for both classical and quantum gravity.

%%%%%%%%%%%%%%%%%%
\section{Black hole minisuperspace}
%%%%%%%%%%%%%%%%%%

We consider the class of static spherically symmetric spacetimes, corresponding to the (rotationless) black hole mini-superspace model.
One can consider at the same time the interior and exterior regions of the black hole by choosing a null gauge
%\footnote{This is a different choice with respect to the literature \cite{Geiller:2020xze, BenAchour:2023dgj}, but perfectly consistent with it, thanks to the invariance of the minisuperspace of the choice of coordinates on the radial slices \cite{Sartini:2022ecp}},
%
regular across the horizon\footnotemark{}:
\be
\label{line}
\de s^2 = 2N(r)\de r \de u + \f{\Vtwo(r)}{2\Vone(r)} \de u^2 + \ls^2 \Vone(r) \de \Omega^2
\ee
with three independent free functions $N, \Vone$ and $\Vtwo$. The length unit $\ls$ allows to keep $\Vone$ and $\Vtwo$ dimensionless and sets the scale for the curvature of the spherical sections $\de \Omega^2=\de \theta^2 + \sin^2 \theta \de \phi^2$.
\footnotetext{
In order to compare with other spherically-symmetric metric ansatz for black hole or compact objects in astrophysics, it is useful to write this line element in terms of $(t,r)$ coordinates, as:
\be
\rd s^{2}=\f{V_{2}}{2V_{1}}\rd t^{2}-N(r)^{2}\f{2V_{1}}{V_{2}}\rd r^{2}+\ell_{s}^{2}V_{1}\rd\Omega^{2}\,,
\nn
\ee
where the time and null coordinates are related by $u=t+r^{*}$ in terms of the tortoise coordinate defined as:
\be
\rd r^{*}=-2N\f{V_{1}}{V_{2}}\rd r
\,.\nn
\ee
Assuming that $V_{1}\ge 0$, we see that the black hole horizon is located at $V_{2}$, with the exterior region for $V_{2}<0$ and the interior region for $V_{2}>0$. The central singularity is located at the root of $V_{1}$. The standard Schwarzschild metric thus corresponds to:
\be
N=1
\,,\quad
\ell_{s}^{2}V_{1}=r^{2}
\,,\quad
\f{V_{2}}{2V_{1}}=-\left(1-\f{2M}{r}\right)
\,.\nn
\ee
}

The black hole mini-superspace is defined by plugging this metric ansatz in the Einstein--Hilbert action for general relativity. As we review below, solutions are given up to symmetries and gauge fixing by the Schwarzschild metric, as expected, in the Eddington--Finkelstein coordinates:
\bg
\label{BHmetric}
\de s^2= 2\de R \de U -\left (1-\f{2M}{R}\right )\de U^2+R^2 \de \Omega^2\,,
\eg
where $M$ denotes the mass of the black hole solution. The black hole mini-superspace thus describes the phase space of Schwarzschild black hole metrics with arbitrary mass and their spherically-symmetric fluctuations.
In some sense, one can interpret the metric ansatz \eqref{line}, with arbitrary components $V_{1}$ and $V_{2}$, as an off-shell black hole, i.e. before imposing the Einstein equations ( or suitably modified Einstein equations).
Below, we review the definition of the mini-superspace, its action, dynamics and symmetries.

\medskip

Evaluating the Einstein-Hilbert action on the line element ansatz \eqref{line} reduces general relativity to a mechanical model and gives the following reduced action, similarly to \cite{Geiller:2020xze, Sartini:2022ecp},
\be
\label{action}
\cS = \f{\cV_0}{\lp^2}  \int \de r\left [\f{ V'_{1}(\Vtwo V'_{1} -2 \Vone V'_{2})}{2N \Vone^2} +\f{4N}{\ls^2}\right ] 
%
%\cS = \int \de r \,L[\Vone,\Vtwo]
%\,,\quad
%L=\f{\cV_0}{\lp^2} \left [\f{ \Vone'(\Vtwo \Vone' -2 \Vone \Vtwo')}{2N \Vone^2} +\f{4N}{\ls^2}\right ] 
\ee
where the prime denotes the derivative with respect to the radial coordinate $r$.
Units are chosen such that the Planck length is $\lp = \sqrt{G}$, with the Newton constant $G$.
There are two essential remarks. First, we are considering stationary space-times, the metric components do not depend on the time coordinate, the fields only depend on $r$, so that the dynamics is entirely along the radial direction.
Second, the metric ansatz is homogeneous, so the Einstein-Hilbert action has to be evaluated on a finite time interval in order to have a well-defined action principle. Here we restrict the null direction $u$ to a finite range $u\in[u_1,u_2]$, which gives the fiducial volume pre-factor $\cV_0 = \ls^2 (u_2-u_1)/8$ in front of the action.

The radial coordinate plays the role of the evolution parameter, and we can develop the Hamiltonian formulation describing this evolution. Let us compute the conjugate momenta by differentiating the Lagrangian with respect to the radial derivatives $\Vone'$ and $\Vtwo'$,
\begin{align}
P_{1}
&=
\f{\pp L}{\pp V'_{1}}
=
-\f{\cV_0}{\lp^2N}\f{\left(V'_{1}\Vtwo-\Vone V'_{2}\right)}{\Vone^{2}}
\,,\\
P_{2}
&=
\f{\pp L}{\pp V'_{2}}
=
-\f{\cV_0}{\lp^2N}\f{V'_{1}}\Vone
\,,\nn
\end{align}
with the reverse formulas:
\be
V'_{1}=-\f{\lp^2N}{\cV_0}V_{1}P_{2}
\,,\quad
V'_{2}=-\f{\lp^2N}{\cV_0}\left(
V_{1}P_{1}+V_{2}P_{2}\right)
\,.
\ee
We have a four-dimensional phase space. We perform the corresponding Legendre transform and write the action in its Hamiltonian form:
\be
\cS=\int \rd r\,\Big{[}
V'_{1}P_{1}+V'_{2}P_{2}-H
\Big{]}
\,,
\ee
with $H=N\cH$ where
\be
\cH=\cH_{r}-\f{4\cV_0}{\lp^2\ls^{2}}
\,,\,\,\,
\cH_{r}=-\f{\lp^2}{2\cV_0}P_{2}
\left(
2V_{1}P_{1}+V_{2}P_{2}
\right)
\,.
\ee
The metric component $N$ does not have any conjugate momentum. It plays the role of a Lagrange multiplier enforcing the constraint, $\cH=0$ or equivalently $\cH_{r}=4\cV_{0}/\lp^{2}\ls^{2}$, generating gauge reparametrizations of the $r$ coordinate.
It indeed corresponds to the  expression of the generator of radial diffeomorphisms  in full general relativity, evaluated on our spherically symmetric ansatz \eqref{line}, and Einstein equations require it to vanish on-shell \cite{Geiller:2020xze, BenAchour:2023dgj, Sartini:2022ecp}. 
We refer to this constraint as the \textit{scalar constraint}. 
We can describe the evolution in term of a gauge-invariant coordinate $\tau$, defined by taking into account the $N(r)$ factor as $\de \tau = N \de r$. The evolution of a phase space observable $\cO$ is then obtained from its bracket with the scalar constraint $\cH$:
\be
%\dot\cO &:= 
\de_\tau \cO
=
\partial_\tau \cO+\{\cO,\cH\}
= 
\partial_\tau \cO+\{\cO,\cH_{r}\}
\,.
\ee
In the following, we will use the dot notation to refer to derivative with respect to this coordinate, $\dot{\cO}=\de_\tau \cO$.

%%%%%%
%Playing with the residual gauge invariance, we will work with a gauge fixed coordinate such that $\de \tau = \cN \de r$. 
%
%In the Hamiltonian formalism, the evolution of a phase space observable $\cO$ is then obtained from its bracket with
%\be
%\cH  &= -\f{\lp^2}{\cV_0} \f{\Ptwo (2 \Pone \Vone + \Ptwo \Vtwo)}{2} - \f{4\cV_0}{\lp^2 \ls^2}\,,\\
%\dot\cO &:= \de_\tau \cO = \partial_\tau +\{\cO,\cH\}\,.\notag
%\ee
%where $P_i$ represents the canonical momenta of the configuration variables $V_i$. Together they span the four-dimensional phase space of the model. The Hamiltonian $\cH$ is obtained from a Legendre transform of \eqref{action}. Without surprises, it corresponds to the gauge-fixed expression of the generator of radial diffeomorphism and it must vanish on-shell \cite{Geiller:2020xze, BenAchour:2023dgj, Sartini:2022ecp}. 
%
%Because of the presence of a constant potential, all the dynamics are encoded in the kinetic term $\Hr= \cH + 4\cV_0/\lp^2 \ls^2$. The potential is thus a central element in the phase space, meaning that for any function $\cO$ we have
%\be 
%\{\cO,\cH\} = \{\cO,\Hr\}
%\ee

%%%%%%%%%%%%%%%%
\section{Phase space Poincar\'e symmetry}
%\section{Poincar\'e invariance of the mechanics}
%%%%%%%%%%%%%%%%

As shown in a previous paper \cite{Geiller:2020xze}, the black hole mini-superspace admits a symmetry under the three-dimensional Poincar\'e group $\R^{2,1}\rtimes \Spin(2,1)$ (which is the double cover of $\ISO(2,1)$), whose Noether charges allow to fully integrate the dynamics of the model. It is important to keep in mind two essential features of this construction:
\begin{itemize}

\item
 These Poincar\'e symmetry transformations act on the field space, spanned by the metric components $V_{i=1,2}$, and should not be confused with the isometries of the metric \eqref{line}. They are not a priori related to space-time diffeomorphisms, but can instead be understood as Killing vectors on the space of metrics
 \footnote{
 This comes from a geometrization of the field space. Indeed the action \eqref{action} corresponds to the geodesic Lagrangian for a  metric on the space of (reduced) metrics parametrized by $V_{1}$ and $V_{2}$, or super-metric in short,
% equations of motion of \eqref{action} correspond to the geodesic equations on the space of configurations, parametrized by the $V_i$'s, embedded with a metric
\be 
%\label{supermetric}
\de s^2_\rm{field}
=
\f{\cV_0}{ \lp^2}\,\bigg{[}\f\Vtwo{\Vone^2}(\de \Vone)^{2}- \f2\Vone \de \Vone\de \Vtwo \bigg{]}
%\f{\cV_0}{\Vone^2 \lp^2}(\Vtwo\de \Vone-2 \Vone \de \Vtwo )\de \Vone
\,.
\nn
\ee
Symmetries are directly read from the properties of this field space metric, as shown in \cite{Geiller:2022baq, BenAchour:2022fif, Sartini:2022ecp}. Indeed,  a set of charges forming a Shr\"odinger algebra $\left (\sl(2,\R) \oplus \so(1,1)\right ) \loplus \h_2 $ can be built out of the conformal Killing vectors of the space of (reduced) metrics (or super-space). The Poincar\'e generators studied here are part of this algebra: the $\cL$'s are the generators of the $\sl(2,\R)$ subalgebra, while the  $\cT$'s are obtained as quadratic combinations of the  Heisenberg subgroup $\h_{2}$.
% \cite{Geiller:2022baq, BenAchour:2022fif, Sartini:2022ecp}.
}.

\item
These are physical symmetries and not gauge symmetries. They act non-trivially on the set of physical trajectories of the system.

\end{itemize}
Let us review the Poincar\'e symmetry transformations in this section, their action on the metric, their Noether charges and how they allow to integrate the equations of motion for $V_{1}$ and $V_{2}$.

The $\Spin(2,1)\sim\SL(2,\R)$ sector of the Poincar\'e group consists in conformal  transformations, which act on the coordinate $\tau$ by M\"obius transformations, while the metric components $V_i$ are fields conformal with weight one: 
%\be
%\tau \mapsto \ttau=f(\tau) = \f{a \tau + b}{c \tau +d}
%\,,\quad
%V_i(\tau) \mapsto \tV_{i}(\ttau)= \dot f V_i\,.
%\ee
\be
{ \everymath={\displaystyle}
\left|
\begin{array}{lcl}
\tau &\mapsto&\ttau=f(\tau) = \f{a \tau + b}{c \tau +d}
\,,\vspace*{1mm}\\
V_i(\tau) &\mapsto &\tV_{i}(\ttau)= \dot f(\tau) V_i(\tau)\,,
\end{array}
\right.}
\ee
where $a,b,c,d\in\R^4$ with $ad-bc=1$.
The abelian sector $\R^{2,1}$ leaves the $\tau$ coordinate invariant, and acts as translations on the metric component $V_{2}$,
\be
\left|
\begin{array}{lcl}
\Vone&\mapsto&\Vone
\,,
\vspace*{1mm}\\
\Vtwo &\mapsto& \Vtwo+ g \dot \Vone - \dot g \Vone\,,
\end{array}
\right.
\ee
for a second-degree polynomial $g(\tau)$. Since $g$ is at most quadratic in $\tau$, this indeed defines a linear space  of dimension 3.
A direct computation allows to check that these are indeed symmetries of the reduced action \eqref{action}.

One can in fact consider arbitrary functional parameters $f$ and $g$ for these transformations defined above. This extend the 3D Poincar\'e group to the 3D BMS group $\mathrm{BMS}_{3}$.  However, these are not symmetries of the theory in general. in fact, they generate interesting extra potential terms in the Lagrangian, and provide non-trivial maps between physically different theories, as explored in \cite{Geiller:2021jmg,Sartini:2021ktb,Sartini:2022ecp}.

The six symmetry transformations lead to six conserved charges, following Noether theorem. The $\SL(2,\R)$ sector gives a first set of three constants of motion:
\begin{align}
\cL_{-}&=-\cH_{r}
\,,\label{eq:Lcharges}\\
\cL_{0}&=-C-\tau \cH_{r}
\,,\nn\\
\cL_{+}&=\f{2\cV_{0}}{\lp^{2}}V_{2}-2 \tau C-\tau^{2}\cH_{r}
\,,
\nn
\end{align}
while the abelian translation sector gives another set of three constants of motion,
\begin{align}
\cT_{-}&=\f{\lp^{2}}{\cV_{0}}A
\,,\label{eq:Tcharges}\\
\cT_{0}&=V_{1}P_{2}+\tau\f{\lp^{2}}{\cV_{0}}A
\,,\nn\\
\cT_{+}&=\f{2\cV_{0}}{\lp^{2}}V_{1}+2\tau V_{1}P_{2}+\tau^{2}\f{\lp^{2}}{\cV_{0}}A
\,,
\nn
\end{align}
where we have written:
\be
\label{CA_def}
C=-\Pone \Vone-\Ptwo \Vtwo\quad \rm{and}\quad  A= \f{\Ptwo^2 \Vone}{2}
\,. 
\ee
$C$ is (minus) the generator of dilatations on the phase space $(V_{i},P_{i})$. These six conserved charges form a $\iso(1,2)$ Poincar\'e algebra, consistently with Noether theorem,
\begin{align}
\label{Iso21}
&\{\cL_{0},\cL_{\pm}\}=\mp\cL_{\pm}
\,,\quad
\{\cL_{+},\cL_{-}\}=2\cL_{0}\,,
\\
&\{\cL_{0},\cT_{\pm}\}=\mp\cT_{\pm}
\,,\quad
\{\cL_{+},\cT_{-}\}=\{\cT_{+},\cL_{-}\}=2\cT_{0}
\,,\nn
\end{align}
with $\{\cT_{a},\cT_{b}\}$ and the remaining Poisson brackets all vanishing.
A neat way to repackage those charges is to write them as:
\begin{align}
\cL_n &= \df{\cV_0}{\lp^2} \Vtwo \ddot \eta -C \dot \eta -\Hr \eta \, \\
\cT_n &= \df{\cV_0}{\lp^2} \Vone \ddot \eta+ \Ptwo \Vone \dot \eta + \df{\lp^2}{\cV_0} A \eta \,
\end{align}
in terms of the parameter function $\eta = \tau^{n+1}$ with $n\in\Z$. These observables form a closed algebra under the Poisson bracket:
\be
%\label{Iso21}
&\{\cL_n,\cL_m \} = (n-m)\cL_{n+m}\,,\q
\{\cT_n,\cT_m \} = 0\,,\\
&\{\cT_n,\cL_m \} = (n-m)\cT_{n+m}\,.\nn
\ee
The Poincar\'e charges corresponds to the case $n=0,\pm 1$, which form a closed sub-algebra. The other observables are not constants of motion, but form the BMS${}_{3}$ algebra uncovered and discussed in \cite{Geiller:2021jmg,Sartini:2021ktb,Sartini:2022ecp}.

Having a four-dimensional phase space, the six Poincar\'e charges can not be independent and must be redundant. In fact, 
%they satisfy two algebraic conditions
the two Poincar\'e Casimirs vanish:
\begin{align}
&\kC_{1}=\vec{\cT}^{2}= \cT_{0}^{2}-\cT_{-}\cT_{+}=0
\,,\\
&\kC_{2}=2\vec{\cL}\cdot\vec{\cT}
=
2\cL_{0}\cT_{0}-\cT_{+}\cL_{-}-\cT_{-}\cL_{+}=0
\,.\nn
\end{align}
The first Casimir corresponds to the mass of the Poincar\'e representation, while the second Casimir gives its spin. This means that the black hole mini-superspace carries a scalar (i.e. zero spin) and massless representation of the Poincar\'e group.
Counting constants of motion and degrees of freedom, we thus have four a priori independent constants of motion in a four-dimensional phase space, implying that the Poincar\'e charges should allow to fully integrate the equations of motion of the system.

Indeed, the Poincar\'e conserved charges \eqref{eq:Lcharges} and \eqref{eq:Tcharges} are actually the initial conditions for the two metric components $V_{1}$, $V_{2}$, their velocities and their accelerations. One can actually inverse the definition of those charges and get the explicit trajectories for $V_{1}$ and $V_{2}$ with the Poincar\'e charges playing the role of integration constants:
\be
\label{integration}
\begin{array}{rl}
\Vone &= \df{\lp^2}{2 \cV_0}(\cT_+ -2 \tau \cT_0 +\tau^2 \cT_-)\,,\\[9pt]
\Vtwo &= \df{\lp^2}{2 \cV_0}(\cL_+ -2 \tau \cL_0 +\tau^2 \cL_-)\,,\\
\end{array}
\ee
where the $\cL$'s and $\cT$'s are constants of motion.
A direct computation allows to check that these are indeed solutions of the equations of motion and amount to exponentiating the flow of the Hamiltonian $\exp\{\cdot,\tau\Hr\}$ on the phase space.

%The presence of this algebra exhibits the integrability of the geodesic motion on the field space. To solve the dynamics of the mechanical model we need in principle four integration constants. The charges above give all of them and are moreover redundant, in the sense that they are not all independent. There are two combinations of them which vanish, corresponding to the Casimir operators of the $\iso(2,1)$ algebra. This also means that the black hole minisuperspace carries a scalar (zero spin) massless representation of the Poincar\'e group. 
%
%Inverting the relationship between the $V_i$'s and the charges, or equivalently exponentiating the time evolution with $\exp\{\bullet,\tau\Hr\}$ we can obtain the classical trajectories. The presence of charges linear in $V_i$ and at most quadratic in $\tau$ means that the solutions for the fields are also quadratic in the gauge fixed radial coordinate. We have indeed
%\be
%\label{integration}
%\begin{array}{rl}
%\Vone &= \df{\lp^2}{2 \cV_0}(\cT_+ -2 \tau \cT_0 +\tau^2 \cT_-)\,,\\[9pt]
%\Vtwo &= \df{\lp^2}{2 \cV_0}(\cL_+ -2 \tau \cL_0 +\tau^2 \cL_-)\,,\\
%\end{array}
%\ee

%using the vanishing of the Poincar\'e Casimir \eqref{casimir} and
The fact that the first Poincar\'e Casimir vanishes, $\kC_{1}= \cT_{0}^{2}-\cT_{-}\cT_{+}=0$, means that the discriminant of the quadratic $(\cT_+ -2 \tau \cT_0 +\tau^2 \cT_-)$ vanishes and that it admits a double root.
Then, keeping in mind that the scalar constraint fixes the value of the Hamiltonian $H_r=-\cL_-=4\cV_0/\lp^2 \ls^2$, the trajectories can be written as in \cite{Geiller:2020xze} as
\be 
\label{solutions}
\begin{array}{rl}
\Vone(\tau) &= \df{A \lp^4 (\tau-\tau_0)^2}{2 \cV_0^2}\,,\\[9pt]
\Vtwo(\tau) &= \df{B\lp^2(\tau-\tau_0)}{\cV_0} - 2\df{(\tau-\tau_0)^2}{	\ls^2}\,.
\end{array}
\ee
The constant of integration $A = \cV_0 \cT_-/\lp^2$ was already introduced earlier in \eqref{CA_def}. The second constant of integration $B= V_{1}P_{1}$ gives the Casimir operator of the $\sl(2,\R)$ subalgebra spanned by the $\cL$'s, explicitly $B^2=\kC_{\sl}= \cL_0^2-\cL_+ \cL_-$.
%is a constant of motion. It is simply related to the Poincar\'e charges, as it gives the Casimir operator of the $\sl(2,\R)$ subalgebra spanned by the $\cL$'s, explicitly $B^2=\kC_{\sl}= \cL_0^2-\cL_+ \cL_-$.
%
Finally, the shift $\tau_0=\cT_0/\cT_-$ gives the location of the singularity. Indeed, at $\tau=\tau_{0}$, the metric component $g_{uu}\propto V_{2}/V_{1}$ diverges. Remember that $\tau$ is the proper coordinate in the radial direction. 

Actually, a change of variables allows to put the singularity back to its usual location at vanishing radius,
\bg
\label{extra_change}
\tau -\tau_0 = \f{\sqrt{2}\cV_0}{\sqrt{A}\ls \lp^2} R \,,\q u=\f{\sqrt{A}\ls \lp^2s}{\sqrt{2}\cV_0} U\,,
\eg
and recover the Schwarzschild metric in the Eddington--Finkelstein coordinates:
\bg
\de s^2= 2\de R \de U -\left (1-\f{2M}{R}\right )\de U^2+R^2 \de \Omega^2\,,
\eg
where the mass is now expressed in terms of the Poincar\'e charges and reads:
\be
M=\f{\sqrt{A} \ls^3\lp^4 B}{4\sqrt{2}\cV_0^2}
= \f{\ls^3\lp^3}{4\sqrt{2}\cV_0^{3/2}} \sqrt{\cT_- \kC_{\sl}}
\,.
\ee
%The mass, as phase space function, is also proportional to the square root of the $\sl(2,\R)$ Casimir, but with an additional, rotational breaking factor, depending on the translation $\cT_-$
Out of the four constants of motion, the value of the Hamiltonian $\cH_{r}$ is fixed in terms of the fiducial scales $\ls$ and $\cV_{0}$, while $A$ and $\tau_{0}$ appear to be gauged out by reparametrization of the coordinates $r$ and $u$. Finally, only the mass $M$ seems to be physical and remains in the final solution metric.
%Out of the four initial conditions, only this combination seems to be physical, as $\cA$ and $\tau_0$ are somehow gauged out by the additional spacetime reparametrization, corresponding respectively to the area measure on the two-sphere (not its total physical area) and to the initial radial coordinate at the singularity. 

%The fourth initial condition is not completely free. It is fixed by the scalar constraint and it explicitly depends on the fiducial sizes $\cV_0$ and $\ls$ \cite{Geiller:2022baq, Sartini:2022ecp}.

%{\bf  MORE DETAILS + Plots, singularity at $V_{1}=0$ and horizon at $V_{2}=0$}

\begin{figure}[h!]
\includegraphics[width=75mm]{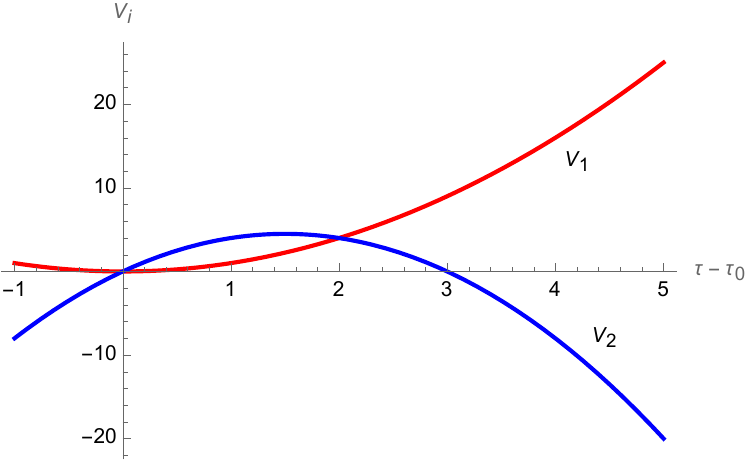}
\caption{Plots of the metric components $V_1$ (in red) and $V_2$ (in blue) in terms of the proper radial coordinate $(\tau-\tau_{0})$ for initial coordinate $\tau_{0}=0$, fiducial length $\ls=1$, radial expansion rate $\gamma=1$, and mass $2\gamma M=3$. The field $V_{1}$ gives the area of the 2-spheres. It is always positive, and its single root $\tau=\tau_{0}$ is the singularity. The standard black hole space-time is usually chosen as covering $\tau>\tau_{0}$. The field $V_{2}$ also vanishes at the singularity, but admits a second root at $\tau=\tau_{0}+2\gamma M$, which gives the mass of the black hole. The region with $V_{2}>0$ is the interior of the black hole where the radial coordinate is time-like, while $V_{2}<0$ is the exterior of the black hole where the radial coordinate is space-like.}
\label{fig:plotV1V2}
\end{figure}

What's important to remember, for the physical interpretation of the mini-superspace, is that the $V_{1}$ field remains always positive but its zero  $V_{1}=0$ is a singularity, thus located at $\tau-\tau_{0}=0$. On the other hand, the $V_{2}$ field can change sign and its zero  $V_{2}=0$  signals the horizon, located at $R=2M$, or equivalently $\tau-\tau_{0}=B\ls^{2}\lp^{2}/2\cV_{0}$, as illustrated on the fig.\ref{fig:plotV1V2}.
To make notations easier to read, we call $\gamma$ the proportionality factor between $R$ and $(\tau-\tau_{0})$, It is equal to $1/\sqrt{A}$ up to dimension factors (depending on $\ls$, $\lp$ and $\cV_{0}$). Then the two metric components read more simply,
\begin{align}
\label{originaltraj}
V_{1}(\tau)&=\f{(\tau-\tau_{0})^{2}}{\gamma^{2}\ls^{2}}
\,,\\
V_{2}(\tau)&=\f{-2(\tau-\tau_{0})}{\ls^{2}}\Big{[}
(\tau-\tau_{0})-2\gamma M
\Big{]}
\,,\nn
\end{align}
keeping in mind that both $V_{1}$ and $V_{2}$ are dimensionless fields.

Avoiding the singularity means finding a way to shift $V_{1}$ to strictly positive values. A direct way to do so is to shift the first Poincar\'e Casimir $\kC_{1}= \vec{\cT}^{2}$, from its 0 value to a negative value, i.e. use a Poincar\'e representation with negative squared mass. It is actually possible to add suitable potential terms of the Hamiltonian in order to shift the two Poincar\'e Casimirs, as we show in section \ref{sec:general}. Although this should definitely be investigated, we could like to focus here on another route, by deforming the Poincar\'e algebra while keeping vanishing Casimirs.

%%%%%%%%%%%%%%%%%%
\section{$\lambda$-deformation}
%\section{Deformation and doubled cosmology}
\label{sec:defo}
%%%%%%%%%%%%%%%%%%%%

We have shown that the black hole mini-superspace admits a Poincar\'e symmetry under $\R^{2,1}\rtimes \Spin(2,1)$ transformations, generated by conserved charges forming a $\iso(2,1)$ Lie algebra. This symmetry algebra is not semi-simple. This usually happens when the considered system is a (degenerate) limit case of a more general model admitting a semi-simple symmetry algebra. For instance, in the context of general relativity, the typical example is the de Sitter isometry group $\SO(d,1)$ or anti-de-Sitter isometry group $\SO(d-1,2)$ leading to the ``degenerate'' Poincar\'e isometry group $\ISO(d-1,1)$ in the limit of a vanishing cosmological constant $\Lambda\to0$.
%
%A natural question is whether we can deform the symmetry algebra to get rid of the degeneracy of the algebra that is not semisimple. The presence of the abelian subalgebra suggests the presence of a deformation parameter, that is here set to zero, giving the degenerate sector of a more general theory. An example of such a mechanism is the double copy of Virasoro in (A)dS-3D gravity that degenerates in the BMS algebra when the cosmological constant goes to zero \cite{Barnich:2012aw}.   
%
Following the same logic, we naturally investigate if the present Poincar\'e symmetry algebra $\iso(2,1)$ could get ``regularized'' to a $\so(3,1)$ or $\so(2,2)$ symmetry algebra, which would drive a generalized (or deformed) black hole mini-superspace.

We will see that it is indeed possible and we will introduce below a $\lambda$-deformation of the phase space of the black hole mini-superspace. The deformation parameter $\lambda$ has a priori no relation whatsoever with the cosmological constant $\Lambda$. The interested reader can actually find details on the mini-superspace of dS or AdS Schwarszchild black holes in \cite{Achour:2021dtj,BenAchour:2022fif,BenAchour:2023dgj}.
Here we will show that, on the one hand, the $\lambda$-deformed black hole phase space can be intriguingly written as a superposition of two copies of the FLRW cosmology phase space, and on the other hand, it leads to regularized black hole metrics with the central singularity replaced by a bounce, similarly to Big Bounce scenarios for regularized FLRW cosmologies (e.g. in loop quantum cosmology \cite{Ashtekar:2008ay,Wilson-Ewing:2012lmx,Linsefors:2013cd} and related approaches \cite{BenAchour:2018jwq}).
Then, at $\lambda=0$, one is back to the standard Schwarzschild black hole mini-superspace with its Poincar\'e phase space symmetry and its solutions with singular metrics.

\medskip

So we would like to understand whether it is possible or not to deform the $\iso(2,1)$ algebra of conserved charges \eqref{Iso21} by introducing a parameter $\defo$ such that the translation charges $\cT$'s seek to be abelian and satisfy the following Poisson brackets,
\be
\label{So_lamb}
\begin{array}{rl}
\{\cL^{\defo}_n,\cL^{\defo}_m \} &= (n-m)\cL^{\defo}_{n+m}\,,\\[4pt]
\{\cT^{\defo}_n,\cT^{\defo}_m \} &= \defo(m-n) \cL^\defo_{n+m}\,,\\[4pt]
\{\cT^{\defo}_n,\cL^{\defo}_m \} &= (n-m)\cT^{\defo}_{n+m}\,,
\end{array}
\ee
giving the Lie algebra $\so(3,1)\sim\sl(2,\C)$ when $\defo>0$, or the Lie algebra $\so(2,2)\sim\sl(2,\R)\times \sl(2,\R)$ when $\defo<0$.
%
%This is equivalent to two copies of $\sl(2,\R)$ if $\defo$ is negative or to the 3+1 Lorentz algebra $\so(3,1)$ when $\defo>0$. 

More precisely, we are looking for a set of charges, forming a closed algebra, such that  the highest-order charges $\cL^\defo_+$ and $\cT^\defo_+$ are still proportional to the fields $\Vi$. The dynamics will still be generated by the charge $\cL_-^{\defo}$, so that the trajectories will still be quadratic in the evolution coordinate $\tau$.

%We want to study the cases when this algebra emerges from a deformation of the dynamics. We ask that the evolution in $\tau$ is always generated by $\cL_0^{\defo}$ and that the highest-order charges $\cL^\defo_+$ and $\cT^\defo_+$ are proportional to the fields $\Vi$. This is equivalent to the requirement that the dynamics is still encoded in this algebra, and the equations \eqref{integration} still hold. In other words, the trajectories are always quadratic in the coordinate $\tau$. 
%
%However, when we have mapped the equation \eqref{integration}, depending on six parameters into the physical solutions \eqref{solutions}, we made use of the vanishing of the Casimirs to go back into the four-parameter family. Now, if we deform the algebra, the new Casimirs still vanish, but we change their functional form in terms of the charges, that in turn represent the derivatives of $V_i$. This causes different dynamics to appear, in particular when we will reconstruct a metric out of the solutions. We will detail the consequences of the deformation on the geometry in the last section.

Searching systematically for possible Hamiltonians leads to a multi-parameter family, which we discuss in details in section \ref{sec:general}.
Out of those, Hamiltonians leading to a deformed symmetry group are parametrized by a single parameter $\lambda$. They are simply given by adding a single correction term:
%Out of them, the physically-relevant deformations are parametrized by a single parameter $\lambda$ and are simply given by adding a term to the Hamiltonian:
%The simplest deformation that achieves the goal presented in the previous paragraph is by a deformation of the Hamiltonian (see Appendix \ref{appA} for the most general deformation) as
\be
\label{Hlambda}
\cH^\defo_{r}&= -\cL_-^\defo = \cH_{r} + \defo \f{\lp^2}{\cV_0} \f{\Pone^2 \Vtwo}{2}
\\
&=-\f{\lp^2}{\cV_0} \f{\Ptwo (2 \Pone \Vone + \Ptwo \Vtwo)-\defo \Pone^2 \Vtwo}{2}\,,\notag
\ee
% This is equivalent to a deformation of the supermetric as
%\be
%\label{supermetric_l}
%\de s^2_\rm{mini} = \f{\cV_0(\Vtwo\de \Vone^2-2 \Vone \de \Vone \de \Vtwo -\defo \Vtwo \de \Vtwo^2) }{\lp^2(\Vone^2+\defo \Vtwo^2)}\,.
%\ee
with the corresponding Lagrangian given by 
%\be
%\label{lagrangian_l}
%L^\defo_\rm{BH} = \f{\cV_0}{\lp^2}\left [\f{\Vtwo \dot \Vone^2-2 \Vone \dot \Vone \dot \Vtwo -\defo \Vtwo \dot \Vtwo^2 }{2(\Vone^2+\defo \Vtwo^2)}- \f{8}{\ls^2}\right ]\,.
%\ee
\be
\label{actionlambda}
\cS^{\defo} = \f{\cV_0}{\lp^2} 
\int \de r
\left[\f{V_{2} V'_{1}{}^{2} -2 V_{1}V'_{1} V'_{2}-\lambda V_{2}V'_{2}{}^{2}}{2N (V_{1}^{2}+\lambda V_{2}^{2})} +\f{4N}{\ls^2}\right] 
\,,\nn
\ee
where the last term in $N$ fixes the value of the Hamiltonian to a non-vanishing constant, $\cH^\defo_{r}=4\cV_{0}/\lp^{2}\ls^{2}$.

Iteratively computing the Poisson brackets between the Hamiltonian and the fields $V_i$, we obtain the full set of  charges:
\begin{align}
\cL_n^\defo &= \df{\cV_0}{\lp^2} \Vtwo \ddot \eta -C \dot \eta -\cH^\defo_{r} \eta
\\
\cT_n^\defo &= \df{\cV_0}{\lp^2} \Vone \ddot \eta+ (\Ptwo \Vone-\defo \Pone \Vtwo) \dot \eta + \df{\lp^2}{\cV_0} A^\defo \eta
\nn
\end{align}
where $\eta(\tau)=\eta_{-}+\eta_{0}\tau+\eta_{+}\tau^{2}$ is a second-degree polynomial in the evolution coordinate as in the undeformed case. $C$ is unmodified while the charge $A$ acquires $\lambda$-corrections:
\begin{align}
C^{\defo} &:= C =-V_{1}P_{1}-V_{2}P_{2} \\
A^\defo &:=\f{\Ptwo^2 \Vone}{2}-\f{\defo}{2}  \Pone^2 \Vone- \defo  \Pone \Ptwo \Vtwo
\,.\nn
\end{align}
These charges $\vec{\cL}_n^\defo$ and $\vec{\cT}_n^\defo$ are conserved by construction along the Hamiltonian flow generated by $\cH^\defo_{r}$.
Setting the deformation parameter to zero, $\defo= 0$, gives back the Poincar\'e algebra $\iso(2,1)$ of the black hole mini-superspace reviewed in  the previous section. For a non-vanishing deformation parameter $\defo$, we now have a modified black hole mini-superspace, driven by a $\so(3,1)$ Lie algebra for a positive deformation parameter $\lambda>0$ or a $\so(2,2)$ Lie algebra for  $\lambda<0$.

The deformed algebra \eqref{So_lamb} still has two Casimirs. The translation sector is not abelian anymore, so that $\cT^{2}$ is not a Casimir anymore, but needs to be modified:
%\begin{align}
%\label{casimir}
%\kC_1 &=  (\cT^\defo_0)^2- \cT^\defo_{-1} \cT^\defo_{1}-\defo \left ((\cL^\defo_0)^2 - \cL^\defo_{1} \cL^\defo_{-1}\right )\,,\\
%\kC_2 &= 2\cL^\defo_0 \cT^\defo_0- \cL^\defo_{1} \cT^\defo_{-1}- \cL^\defo_{-1} \cT^\defo_{1}\,,\nn
%\end{align}
%\begin{align}
%&\kC_{1}=(\vec{\cT}^\defo)^{2}-\lambda (\vec{\cL}^\defo)^{2}
%%= \cT_{0}^{2}-\cT_{-}\cT_{+}
%\,,\\
%&\kC_{2}=2\vec{\cL}^\defo\cdot\vec{\cT}^\defo
%\,.\nn
%\end{align}
\be
\kC_{1}^\defo=(\vec{\cT}^\defo)^{2}-\lambda (\vec{\cL}^\defo)^{2}
\,,\quad
\kC_{2}^\defo=2\vec{\cL}^\defo\cdot\vec{\cT}^\defo
\,.
\ee
These Casimirs have vanishing Poisson brackets with the charges $\cT^\defo$'s and $\cL^\defo$'s.
Computing the norms and scalar product,
\be
(\vec{\cL}^\defo)^{2}=V_{1}^{2}+\lambda V_{2}^{2}
\,,\quad
(\vec{\cT}^\defo)^{2}=\lambda\,(\vec{\cL}^\defo)^{2}
\,,\quad
\vec{\cL}^\defo\cdot\vec{\cT}^\defo=0
\,,\nn
\ee
we find that both Casimirs vanish as in the undeformed case:
\be
\kC_{1}^\defo=\kC_{2}^\defo=0\,,
\ee
meaning that the system carries a spinless and massless representations of the Lorentz algebra $\so(2,2)$ or $\so(3,1)$ depending on the sign of the deformation parameter $\lambda$.
%
%both vanishing when we write the generators on the phase space parametrized by $V_i$, $P_i$. When $\defo=0$ the algebra becomes the (2+1) Poincar\'e one, and the two functions above are respectively the mass and the spin of the representation.

When $\defo>0$,  we get the Lorentz algebra $\so(3,1)$, which can be re-packaged in terms of 3d rotations and 3d boosts.
When $\defo<0$, the algebra $\so(2,2)$ can be decomposed into two commuting copies of $\sl(2,\R)$,
\bg
\label{kpm}
\cK^\pm_n := \f{\cL_n^\defo \pm  \cT_n^\defo/\sqrt{-\defo}}{2} \,,\\
\{\cK^\pm_n,\cK^-_m \} =(n-m) \cK_{n+m}^\pm \,, \q\{\cK^+_n,\cK^-_m \} = 0 \notag\,.
\eg
In this case, the two Casimirs $\kC_{1}^\defo$ and $\kC_{2}^\defo$ are linear combinations of the two $\sl_{2}$ Casimirs,
\be
\kC_\sl^{\pm} = (\cK^\pm_0)^2- \cK^\pm_{-} \cK^\pm_{+}
\,,\,\,\,
\left|
\begin{array}{l}
\kC_{1}^\defo=
-2{\lambda}(\kC_\sl^{+}+\kC_\sl^{-})
\,,\\[2mm]
\kC_{2}^\defo=
2\sqrt{-\lambda}(\kC_\sl^{+}-\kC_\sl^{-})
\,,
\end{array}
\right.
\ee
implying that the $\sl_{2}$ Casimirs must both vanish, $\kC_\sl^{\pm}=0$.

As we will explain in the next section, each $\sl(2,\R)$ sector  can be mapped onto a FLRW cosmology, so that the deformed black hole mini-superspace for $\lambda<0$ can surprisingly be understood as a superposition of two FLRW cosmologies.

%%%%%%%%%%%%%%%%%%
\section{Doubled cosmology}
\label{sec:doubled}
%%%%%%%%%%%%%%%%%%%%

Let us focus on the case of a negative deformation parameter $\lambda$, when the symmetry Lie algebra $\so(2,2)$ splits as the direct sum of two copies of the $\sl(2,\R)\sim \so(2,1)$ Lie algebra.
Such a $\sl(2,\R)$ symmetry has already been encountered in the context of gravitational mini-superspaces for FRW cosmologies, as originally shown in a series of works \cite{Pioline:2002qz,BenAchour:2019ywl,BenAchour:2019ufa,BenAchour:2020xif,BenAchour:2020ewm}.
Indeed, let us consider  homogeneous isotropic geometries with flat spatial slices, described by the metric ansatz,
\be
\rd s^{2}_{FRW}=-N(t)^{2}\rd t^{2}+a(t)^{2}\delta_{ij}\rd x^{i}\rd x^{j}
\,.
\ee
Then the reduced Einstein-Hilbert action for a matter fluid coupled to such geometry reads:
\be
\cS^{(w)}_\rm{FRW} = -\cV_0 \int \de t \left [\frac{ (\rd_{t} v)^2}{2 \lp^2 N v}+\f{12 \pi  N  \varrho_0}{v ^w}\right ]
\,,
\ee
where $v=a^{3}$ is the spatial volume, $\cV_{0}$ is the fiducial volume of a 3d spatial cell over which we integrate the Einstein-Hilbert action, $\varrho_0$ is the fluid energy density (at $a=1$), and $w$ is the standard parameter encoding the equation of state for the matter
\footnote{Using the standard normalization in cosmology, for which at present-day the scale factor is $a=1$, then $\varrho$ represents the energy density of the fluid today. From the continuity equation, i.e. the conservation of energy, and the state equation we get the evolution of the density: $\varrho(t) = \varrho_0/ v(t)^{1+w}$}.
For instance, a perfect fluid is given by $w=0$ while a free massless scalar field corresponds to $w=1$. We focus here on the case of the perfect fluid, thus setting the equation of state parameter to $w=0$,
\be
\label{cosmo_action}
\cS_\rm{FRW} = -\cV_0 \int \de t \left [\frac{ (\rd_{t} v)^2}{2 \lp^2 N v}+{12 \pi  N  \varrho_0}\right ]
\,,
\ee
Performing the canonical analysis, we compute the cougate momentum to the space volume $v$ and the Hamiltonian:
\be
\label{eq:Hcosmo}
p=-\f{\cV_{0}}{\lp^{2}}\f{\rd_{t}v}{Nv}
\,,\quad
H_{cosmo}=N(\cH_{FRW}+12\pi \cV_{0}\varrho_{0})
\,,
\ee
where the scalar constraint now reads,
\be
\cH_{FRW}=-\f{\lp^{2}}{2\cV_{0}}vp^{2}
\,.
\ee
This constraints generate the evolution of the cosmological system in the proper time $\tau$ defined as $\rd \tau=N\rd t$,
\be
\rd_{\tau}\cO=N^{-1}\rd_{t}\cO=\{\cO,\cH_{FRW}\}\,.
\ee
Writing $C_{FRW}=-vp$, we identify  conserved charges, which mimic the $\sl(2,\R)$ sector of our black hole mini-superspace given in equation \eqref{eq:Lcharges}:
\begin{align}
\cK_{-}&=-\cH_{FRW}
\,,\label{eq:cosmocharges}\\
\cK_{0}&=-C_{FRW}-\tau \cH_{FRW}
\,,\nn\\
\cK_{+}&=\f{2\cV_{0}}{\lp^{2}}v-2 \tau C_{FRW}-\tau^{2}\cH_{FRW}
\,,
\nn
\end{align}
which indeed form a $\sl(2,\R)$ Lie algebra,
\be
\label{sl2r}
\{\cK_{0},\cK_{\pm}\}=\mp\cK_{\pm}
\,,\quad
\{\cK_{+},\cK_{-}\}=2\cK_{0}\,.
\ee
Moreover, one can check that the $\sl_{2}$ Casimir vanishes,
\be
\kC_\sl = (\cK_0)^2- \cK_{-} \cK_{+}=0\,.
\ee
These charges are the initial conditions for the volume, its velocity and acceleration, giving the trajectories in proper time:
\be
\label{eq:cosmoevo}
v=\f{\lp^2}{2 \cV_0}(\cK_+ -2 \tau \cK_0 +\tau^2 \cK_-)\,,
\ee
where the charge $\cK_{-}=-\cH_{FRW}\propto vp^{2}$ is fixed to $12\pi \cV_{0}\varrho_{0}$ by the Hamiltonian constraint \eqref{eq:Hcosmo}.
Moreover, the fact that the Casimir vanishes means that $\cK_{0}^{2}=\cK_{-}\cK_{+}$,
%or equivalently $\cK_{0}=\eps\sqrt{\cK_{-}\cK_{+}}$ up to a sign $\eps$,
which implies that the quadratic polynomial in $\tau$ above has a double root:
\be
v=\cK_{-}(\tau-\tau_{0})^{2}
\textrm{ with }
\left|
\begin{array}{l}
\cK_{-} = 12\pi \cV_{0}\varrho_{0}
\,,\vspace*{2mm}\\
\tau_{0}={\cK_{0}/\cK_{-}}
%\tau_{0}=\eps\sqrt{\cK_{+}/\cK_{-}}
\,.
\end{array}
\right.
\ee
In particular, a positive matter density $\varrho_{0}>0$ leads to a positive volume $v>0$, as physically expected.

\medskip

In light of this symmetry analysis, it seems natural to try to reformulate the black hole mini-superspace with negative deformation parameter $\lambda<0$ as two copies of FRW cosmologies.
In fact, the mapping is rather natural. Let us define the following linear combinations of the two black hole metric components $V_{1}$ and $V_{2}$:
\be
\label{vpm}
v_\pm (\tau) = \f{\Vone\pm \Vtwo\sqrt{-\defo}}{2\sqrt{-\defo}}\,,
\quad
\left|
\begin{array}{l}
V_{1}=\sqrt{-\lambda}(v_{+}+v_{-})\,, \vspace*{1mm}\\
V_{2}=(v_{+}-v_{-})
\,.
\end{array}
\right.
\ee
Using these variables, one can recast the deformed black hole mini-superspace action \eqref{Hlambda} as
\begin{align}
\cS^{\defo} &=
%\f{\cV_0}{\lp^2} 
\cV_0
\int \rd r
\left[
\f{(\rd_{r}v_{-})^{2}}{2N\lp^2 v_{-}}
-\f{(\rd_{r}v_{+})^{2}}{2N\lp^2 v_{+}} 
+\f{4N}{\lp^2\ls^2}
\right]
\,,\nn\\[1mm]
&=
\cS_{FRW}[v_{+}]-\cS_{FRW}[v_{-}]
\end{align}
with the matter energy densities related to the fiducial scales of the black hole mini-superspace by
\be
\label{rhodiff}
3\pi(\varrho^{-}_{0}-\varrho^{+}_{0})
=
\f{1}{\lp^2\ls^2}
\,.
\ee
These relations further hold for positive deformation parameter $\lambda>0$. Then the 3d volume $v_{\pm}$ are no longer real, they are complex numbers, conjugate to one another. The idea of working complex metrics might feel awkward, but has recently been revived in the context of path integrals over cosmological metrics and the study of their complex saddle points, see e.g. \cite{Witten:2021nzp,Jonas:2022uqb,Han:2021kll}.

\begin{figure}[h!]
\includegraphics[width=75mm]{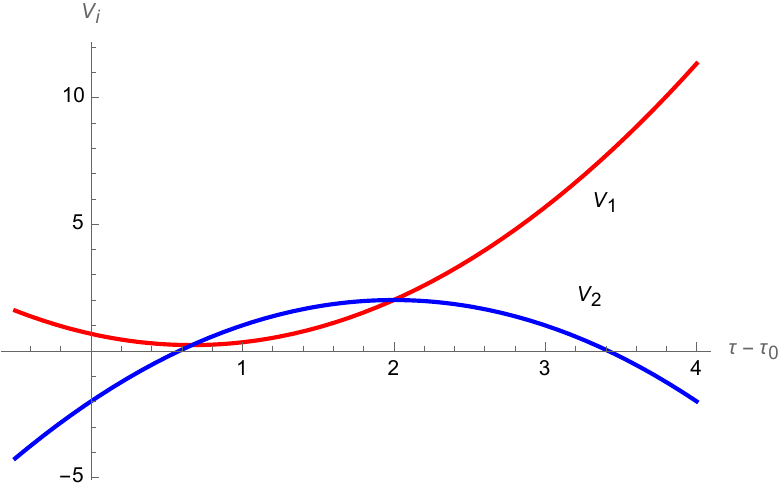}
\caption{Plots of the metric components $V_1$ (in red) and $V_2$ (in blue) in terms of the proper radial coordinate $(\tau-\tau_{0})$ 
constructed from the cosmological volumes $v_{\pm}=12\pi \cV_{0}\varrho_{0}^{\pm}(\tau-\tau_{0}^{\pm})^{2}$, for matter densities $12\pi \cV_{0}\varrho_{0}^{+}=1$ and $12\pi \cV_{0}\varrho_{0}^{-}=2$, and initial cosmological times $\tau_{0}^{+}=0$ and $\tau_{0}^{-}=1$, and deformation parameter $\sqrt{-\lambda}=1/3$.
The field $V_{1}$ remains strictly positive and never vanishes. There is no singularity and the black hole metric bounces back from its minimal area configuration. The phenomenology of the field $V_{2}$ remains unchanged: positive $V_{2}$ is the black hole interior, while negative $V_{2}$ is the black hole exterior. But, since there is no singularity, the black hole metric extends to the whole real line $\tau\in\R$ and we have two exterior regions for the black holes, on both side of the minimal area configuration.}
\label{fig:plotV1V2bis}
\end{figure}

So we have mapped the $\lambda$-modified black hole mini-superspace for spherically-symmetric metric onto a double copy of FRW cosmologies for homogeneous isotropic space-time filled with a perfect fluid. We would like to make two important remarks:
\begin{itemize}

\item We have a superposition of two  FRW cosmologies, but coming with a different sign in the action, which can be interpreted as a flipped direction for the evolution in time.

\item One should keep in mind that we are studying the evolution of the black hole metric components along the radial direction, which we have thus mapped onto the evolution of the FRW cosmological metric along the time direction. Let us not forget nonetheless that the radial coordinate becomes time-like inside the black hole, so that the black hole interior region can truly be considered as a superposition of two FRW cosmologies.
%This switch does not create any problem at the mathematical level, but its physical interpretation remains intriguing.

\end{itemize}
Keeping these points in mind, one automatically gets the trajectories, $V_{1}(\tau)$ and $V_{2}(\tau)$, for the black hole metric in the deformed model from the cosmological evolution $v_{\pm}(\tau)$ given above in \eqref{eq:cosmoevo},
\be
v_{\pm}(\tau)
=
12\pi \cV_{0}\varrho_{0}^{\pm}
(\tau-\tau_{0}^{\pm})^{2}\,,
\ee
with positive matter densities $\varrho^{-}_{0}>\varrho^{+}_{0}>0$.
Since the cosmological volumes $v_{\pm}$ both remain positive, the metric component $V_{1}=\sqrt{-\lambda}(v_{+}+v_{-})$ always remains positive and can never vanish, as illustrated on fig.\ref{fig:plotV1V2bis}. This means that there is no singularity: the $\lambda$-deformation of the black hole mini-superspace regularizes the black hole metric and totally avoids the singularity, at least in the case of  a negative deformation parameter $\lambda<0$.  On the other hand, $V_{2}=(v_{+}-v_{-})$ can still vanish and change sign, which allows to identify the interior and exterior regions of the modified black hole space-time.

The original black hole mini-superspace, with vanishing $\lambda\to 0^{-}$, is recovered by taking the infinite matter densities limit, with both $\varrho_{-}$ and $\varrho_{+}$ scaling in $1/\sqrt{-\lambda}$ while respecting the fixed difference equation \eqref{rhodiff}, and merging the two initial cosmological times, $\tau_{0}^{-}=\tau_{0}^{+}+\gamma M\sqrt{-\lambda}\to\tau_{0}^{+}$, where $M$ is the black hole mass. A quick computation of this limit allows to recover the undeformed expected trajectories \eqref{originaltraj}.

To conclude this section, we stress that the singularity avoidance property is a direct consequence of having regularized the symmetry of the black hole mini-superspace from the Poincar\'e algebra $\iso(2,1)$ to the Lorentz algebra $\so(2,2)$. Let us have a closer look at this important feature in the next section.

\section{Singularity regularization}
%\section{Supermetric and singularity regularization}
\label{sec:singul}
%%%%%%%%%%%%%%%%%%%%%%%%

Another method to solve the dynamics of the deformed mini-superspace is to realize that it is merely a non-linear redefinition of the original mini-superspace. Indeed, let us define the variables,
\be
\label{NLchange}
v_1= \f{\Vone +\sqrt{\Vone^2+ \defo \Vtwo^2}}{2}\,,\quad 
v_2 = \Vtwo \,,
\ee
or reversely,
\be
\label{NLchange-inverse}
\Vone= v_1 - \defo \f{v_2^2}{4v_1}
\,,\quad 
 \Vtwo=v_2 \,.
\ee
This trivializes the kinetic term of the deformed action:
\be
\f{V_{2} V'_{1}{}^{2} -2 V_{1}V'_{1} V'_{2}-\lambda V_{2}V'_{2}{}^{2}}{2(V_{1}^{2}+\lambda V_{2}^{2})}
=
\f{v_{2} v'_{1}{}^{2} -2 v_{1}v'_{1} v'_{2}}{2v_{1}^{2}}
\,.
\ee
This means that we have mapped the deformed black hole mini-superspace back onto the original undeformed mini-superspace; in particular, the variables $v_{1}$ and $v_{2}$ will follow the undeformed equations of motion.

This is similar to the approach introduced in \cite{BenAchour:2018jwq} to generate polymerised FRW cosmology (as in loop quantum cosmology) from standard FRW cosmology through non-linear canonical transformations. However, a canonical transformation does not affect the symmetry of the theory, while here our non-linear field redefinition lifts the degeneracy of the Poincar\'e symmetry $\iso(2,1)$ and changes it  to a Lorentz symmetry $\so(2,2)$ or $\so(3,1)$.
Thus, although the trivializing change of variable \eqref{NLchange} given above looks simple, it is not an innocent field redefinition and deeply affects the physics of the system.

Solving the evolution for the new variables $v_{1,2}$ using the results \eqref{solutions} for the undeformed black hole mini-superspaces, we have the following trajectories in proper time:
\be 
\label{original}
\begin{array}{rl}
v_{1}(\tau) &= \df{a \lp^4 (\tau-\tau_0)^2}{2 \cV_0^2}\,,\\[9pt]
v_{2}(\tau) &= \df{b\lp^2(\tau-\tau_0)}{\cV_0} - 2\df{(\tau-\tau_0)^2}{	\ls^2}\,,
\end{array}
\ee
where we have written $a,b$ for the two conserved charges. Remember that the root of $v_{1}$, at  $\tau=\tau_{0}$, is the black hole singularity, while the other root of $v_{2}$ is the horizon. 
On the one hand, since $V_{2}=v_{2}$, the horizon is not affected by the $\lambda$-deformation.
On the other hand, the $V_{1}$ field differs from $v_{1}$ and acquires a $\lambda$-term:
\be
\label{change}
\Vone (\tau) = \f{a (\tau-\tau_0)^2 \lp^4}{2 \cV_0^2}-\f{\defo  \left(b \lp^2 \ls^2-2 (\tau-\tau_0) \cV_0\right)^2}{2 a \lp^4 \ls^4}\,.
\ee
The most interesting feature is that, for a negative deformation parameter $\lambda<0$, the metric component $\Vone= v_1 - \defo v_2^2 /4v_1$ clearly never vanishes and always remains strictly positive.
Intuitively, $V_{1}^{2}$ gives the area of spatial 2-spheres at constant radial distance. Thus the two-spheres never shrink to a point.
This means that  the singularity is avoided, and replaced by a bounce in the interior region as in a black hole to white hole transition (see e.g. \cite{Olmedo:2017lvt,Bodendorfer:2019nvy,Rovelli:2014cta,Han:2023wxg}).
\begin{figure}[h!]
\includegraphics[width=75mm]{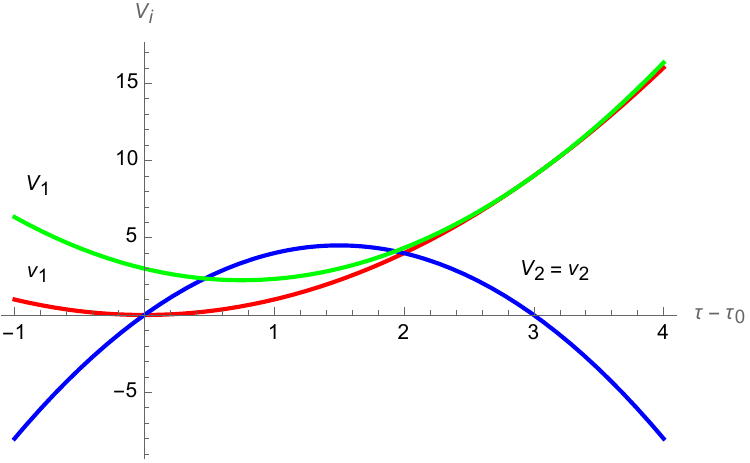}
\caption{Plots of the metric components $v_1$ (in red), $V_1$ (in green) and $V_2=v_{2}$ (in blue) in terms of the proper radial coordinate $(\tau-\tau_{0})$, for a deformation parameter $\lambda=-1/3<0$ and the same constants of motion as the undeformed case in fig.\ref{fig:plotV1V2}.
While the original $v_{1}$ field vanishes at $\tau=\tau_{0}$, signalling the singularity of the classical black hole geometry, the modified trajectory $V_{1}=v_{1}-\lambda v_{2}^{2}/4v_{1}$  remains strictly positive and never vanishes. The black hole metric bounces back from its minimal area configuration and we have a modified black hole solution with no singularity.}
\label{fig:plotV1V2ter}
\end{figure}

%While $\Vtwo$ is unchanged in \eqref{change_field}, we get a deformation of the two-sphere area $\Vone= v_1 - \defo v_2^2 /4v_1$, that implies
%\be
%\Vone (\tau) = \f{A (\tau-\tau_0)^2 \lp^4}{2 \cV_0^2}-\f{\defo  \left(B \lp^2 \ls^2-2 (\tau-\tau_0) \cV_0\right)^2}{2 A \lp^4 \ls^4}\,.
%\ee
%We have the following relationship between $A$ (i.e the second derivative of $v_1$) and $A^\defo$ (the acceleration of $V_1$):
%\be
%A^\defo = A - \f{4\defo   \cV_0^4}{ A \lp^8 \ls^4}\,.
%\ee
%For negatives $\defo$, the deformation corresponds to a singularity resolution, analogous to the one in loop quantum cosmology for the volume of a flat homogeneous universe. A naive illustration of why there is no singularity is given by the fact that the two-sphere never shrinks to a point, being its area $\Vone$ always positive.
This can also be verified by direct computation of the Kretschmann scalar. Performing the same change \eqref{extra_change}. of variables from $(\tau,u)$ to $(R,U)$,  the on-shell line element reads
\bg 
\label{regular_metric}
%\de s^2 = 2 \de R \de U -\f{R (R-2 M)}{F(R)} \de U^2+ F(R) \de \Omega^2
\de s^2
=
2 \de R \de U
%-\f{R^{2}}{F(R)}\left(1-\f{2M}R\right) \de U^2
-\f{R (R-2 M)}{F(R)} \de U^2
+ F(R) \de \Omega^2
\,,
%\\\notag \rm {with}\; 
%\left \{
%\begin{array}{rl}
%F(R)&=(\Delta +1) R^2-4 \Delta  R M+4 \Delta  M^2\\
%\Delta &:= - \defo\df{4 \cV_0^4}{A^2  \lp^8 \ls^4}\\
%\end{array}\right .\,.
\eg
\be
\rm {with} \quad F(R)=R^2- \defo\df{4 \cV_0^4}{A^2  \lp^8 \ls^4}(R-2M)^{2}
\,.
\ee
The Kretschmann scalar blows up if and only if $F(R)$ vanishes and this never happens when $\defo$ is negative, so this is indeed a non-singular space-time metric. One further checks that the Kretschmann scalar goes to 0 at spatial infinity $R\to \infty$, so the solution is asymptotically flat.

%The line element above is obtained by replacing the solutions \eqref{change} into \eqref{line}, and performing the same coordinate change as in the first section \eqref{extra_change}.
%
%In terms of F:
%\be
%\frac{27 R^2 (R-2 M)^2 F'(R)^4+4 F(R)^2 \left(3 R^2 (R-2 M)^2 F''(R)^2+\left(26 R^2-52 R M+20 M^2\right) F'(R)^2+12 R \left(R^2-3 R M+2 M^2\right) F'(R) F''(R)\right)-16 F(R)^3 \left(R (R-2 M) F''(R)+4 (R-M) F'(R)\right)-8 R F(R) (R-2 M) F'(R)^2 \left(4 R (R-2 M) F''(R)+11 (R-M) F'(R)\right)+32 F(R)^4}{4 F(R)^6}
%\ee
%
%with the explicit form of F:
%\be
%\frac{16 M^2 \left(3 \left(\Delta ^2-1\right)^2 R^6-4 \Delta  \left(9 \Delta ^3+5 \Delta ^2-9 \Delta -5\right) R^5 M+4 \Delta  \left(45 \Delta ^3+40 \Delta ^2-4 \Delta -10\right) R^4 M^2-16 \Delta ^2 \left(30 \Delta ^2+30 \Delta +11\right) R^3 M^3+16 \Delta ^2 \left(45 \Delta ^2+40 \Delta +14\right) R^2 M^4-64 \Delta ^3 (9 \Delta +5) R M^5+192 \Delta ^4 M^6\right)}{\left((\Delta +1) R^2-4 \Delta  R M+4 \Delta  M^2\right)^6}
%\ee
%
%For $R\to \infty$ it goes to 0!
%
%The deformation parameter is linked to the cosmological densities. Using the solutions \eqref{change}, the definition of $v_\pm$, the relationship between $\varrho$ and the acceleration of $v_\pm$ and the constraint \eqref{density_negative} we get
%\be
%\label{delta_rho}
%\frac{1}{\sqrt{\Delta}}+\sqrt{\Delta}=6\pi \lp^2 \ls^2 (\varrho_+-\varrho_-) = 2\f{\varrho_--\varrho_+}{\varrho_++\varrho-}\,.
%\ee
%This also allows us to check that one of the volume $(v_-)$ is negative, not only asymptotically, but trough out its whole evolution.

It is crucial to compare this metric to modified Schwarzschild metrics derived in modified gravity theories or in quantum gravity phenomenology, in order to  understand if the scenario considered here has already been realized in the existing literature from another point of view or if it is a brand new mechanism for regular modified black hole metric based on symmetry deformation. 

A look through other quantum gravity scenarios avoiding the singularity, such as in polymerized black holes with bounce black-to-white transitions e.g.\cite{Han:2023wxg,Ashtekar:2005qt,Modesto:2005zm,Boehmer:2007ket,Olmedo:2017lvt,Ashtekar:2018cay,Ashtekar:2018lag,BenAchour:2017ivq,BenAchour:2018khr,Bodendorfer:2019cyv,Bodendorfer:2019nvy,Zhang:2021wex,Geiller:2020xze,Rovelli:2014cta},
reveals that the metric above \eqref{regular_metric} derived for the $\lambda$-deformation is apparently new.
Moreover, although some approaches derive regularized black hole metrics from bouncing cosmology solutions for the black hole interior and/or as superposition of black hole and white hole metrics, none implement a scenario similar to the superposition of expanding and contracting cosmology explicitly derived here.
The ``regularization by symmetry deformation'' scenario developed here can thus be considered as a legitimately new mechanism in quantum gravity phenomenology.

Let us give a few examples.
Keeping in mind how to write spherically-symmetric ansatz in Eddington-Finkelstein coordinates,
\ba
\rd s^{2}
&=&
-\beta(r) \rd t^{2}+\alpha(r)^{-1}\rd r^{2}+\gamma(r)\rd\Omega^{2}
\nn\\
&=&
2N\rd u \rd r +\f{V_{2}}{2V_{1}}\rd u^{2}+\ell_{s}^{2}V_{1}\rd \Omega^{2}
\,,
\ea
where the coordinate $u=t+r^{*}$ is defined in terms of the tortoise coordinate $\rd r^{*}=(\alpha\beta)^{-1/2}\rd r$ and the correspondence between metric components is given by:
\be
%\textrm{with}\quad
V_{1}=\ell_{s}^{-2}\gamma\,,\quad
V_{2}=-2\ell_{s}^{-2}\beta \gamma\,,\quad
N=\sqrt{\beta/\alpha}
\,,
\ee
then the standard polymerized black hole metric ansatz \cite{Modesto:2005zm} is given by
\be
\beta=
\f{(r-r_{+})(r-r_{-})(r+\sqrt{r_{+}r_{-}})^{2}}{r^{4}+a_{0}^{2}}
\,,
\ee
\be
\alpha=\f{(r-r_{+})(r-r_{-})r^{4}}{(r+\sqrt{r_{+}r_{-}})^{2}(r^{4}+a_{0}^{2})}
\,,\quad
\gamma=r^{2}+\f{a_{0}^{2}}{r^{2}}\,,
\ee
with two roots $r_{\pm}$ (like a charged black hole) and a regularization area $a_{0}$ (usually of Planck size),
while the more recent black-hole-white-hole solutions \cite{Rovelli:2014cta,Kelly:2020uwj,Han:2023wxg} is given by
\be
\alpha=\beta=1-\f{2M}r+\f{a_{0}M^{2}}{r^{4}}
\,,\quad
\gamma=r^{2}
\ee
which both do not read the same as \eqref{regular_metric}.
The most similar approach in spirit is the derivation of polymerized-like black holes in \cite{Geiller:2020xze} from a non-linear canonical transformation of the standard Schwarzschild phase space, which leads to (with re-adjusted pre-factors)
\be 
\begin{array}{rl}
V_{1}(\tau) &=
\df{a \lp^4 \tau^2}{2 \cV_0^2}
+\mu_{1}^{2}\df{b^{2}}{2a\tau^{2}}
\,,\\[9pt]
V_{2}(\tau) &= \df{b\lp^2}{\cV_0}\left(\tau+\df{\mu_{2}^{2}}\tau\right) - \df{2}{\ls^2}\left(\tau+\df{\mu_{2}^{2}}\tau\right)^{2}
\,,
\end{array}
\ee
in terms of two Planck-size regularization length scales $\mu_{1,2}$. This can be directly compared to our metric coefficients, \eqref{original} for standard Schwarzschild and \eqref{change} for its $\lambda$-deformation.
However, this construction based on a canonical transformation is designed to preserve all the symmetries of the system, thus to keep the $\iso(2,1)$ symmetry. The $\lambda$-deformation is specifically designed to go beyond this restriction, by deforming the $\iso(2,1)$ symmetry into a $\so(2,2)$ symmetry (in the negative $\lambda$ case).

It would be enlightening to further compare our modified black holes  to other scenarios from quantum gravity, e.g. \cite{Mathur:2005zp,Skenderis:2008qn}
and  semi-classical general relativity, e.g. \cite{Visser:2008rtf,Barcelo:2009tpa,Ho:2019pjr,Kawai:2020rmt}.
This would involve checking the conserved charge algebra and symmetry of the dynamics of those modified black hole mini-superspace.
We postpone such a broad and systematic study to future investigation.

%%%%%%%%%%%%%%%%%%%%
\section{General deformation}
%\label{appA}
\label{sec:general}
%%%%%%%%%%%%%%%%%%%%

In this section, we would like to come back to determining the most deformation of the Hamiltonian for the black hole mini-superspace. We are looking for Hamiltonians for which the two fields $V_{1}$ and $V_{2}$ evolve at most quadraticall in the proper coordinate $\tau$, meaning that their third iterative bracket with the Hamiltonian should vanish. Then we can systematically construct conserved charges, $\vec{\cL}$ and $\vec{\cT}$, out of the first and second derivatives of $V_1$ and $V_{2}$, as in \eqref{integration}. They automatically form a closed Lie algebra.

We find that the most general deformation of the dynamics is achieved by the following family of Hamiltonian, up to a constant shift,
\be
\cH =& -\f{\Ptwo}{2} \left (2\Pone \Vone +\Ptwo \Vtwo\right ) + \f{\defo}{2} \Pone^2 \Vtwo +\cc_3\Ptwo  - f(x)\Pone 
\notag\\
&+ \f{1}{\Vone}\left (\f{ \cc_1 +  x \cc_2 +   \cc_3 f(x) + \f12x f(x)^2}{1 +  x^2 \defo}\right )\,,
\ee
where we have defined $x:= \Vtwo/\Vone$ to make the expression more readable. Corrections to the original black hole mini-superspace Hamiltonian are thus parametrized by three real constants $\lambda, \cc_{i=1..3}\,\in\R$ and an free arbitrary function $f(x)$.

The parameter $\lambda$ is the only one deforming the resulting symmetry algebra, from $\iso(2,1)$ to either $\so(2,2)$ or $\so(3,1)$ depending on its sign. The other parameters deforms the dynamics and trajectories without affecting the Poincar\'e symmetry structure. More precisely, $\cc_{1}$ and $\cc_{2}$ control potential terms, respectively in $1/V_{1}$ and in $V_{2}/V_{1}^{2}$, which  give non-trivial values to the two Casimirs. The two other terms, controlled by the parameters $\cc_{3}$ and $f(x)$, are generated by canonical transformations along respectively $x$ and $P_{2}$. As such, they do not affect the symmetry group of the system.

\medskip

The $\lambda$-deformation has been the focus of the previous sections. Now, in order to understand the effect of  $\cc_{1}$ and $\cc_{2}$, let us switch off all the other deformation parameters, $\lambda=\cc_{3}=f(x)=0$, and consider the Hamiltonian with only the two extra-terms controlled by $\cc_{1}$  and $\cc_{2}$,
\be
\cH^{(\cc_{1},\cc_{2})}
=
\cH^{(0)}+\f{\cc_{1}}{V_{1}}+\f{\cc_{2}V_{2}}{V_{1}^{2}}
\,.
\ee
We are thus adding two potential terms to the Hamiltonian. These terms play the same role as the scalar matter field term in FRW cosmology \cite{BenAchour:2018jwq,BenAchour:2019ywl}.
By computing the iterative Poisson brackets of the metric components $V_{1}$ and $V_{2}$ with this Hamiltonian, we find that the  conserved charges, \eqref{eq:Lcharges} and \eqref{eq:Tcharges}, do not change, except for $\cL_{-}$ which is now minus the deformed Hamiltonian and the constant of motion $A$, which acquires an extra term:
\be
A^{(\cc_{1},\cc_{2})}=
A^{(0)}+\f{\cc_{2}}{V_{1}}
\,.
\ee
A quick calculation shows that the algebra of those charges does not change. It is actually surprising that the addition of potential terms does not break the symmetry of the system. It is still the Poincar\'e $\iso(2,1)$ Lie algebra, but the two Casimirs do not vanish anymore. Their values are directly given by the two deformation parameters:
\begin{align}
&\kC_{1}=\vec{\cT}^{2}=2\cc_{2}
\,,\\
&\kC_{2}=2\vec{\cL}\cdot\vec{\cT}
=2\cc_{1}
\,.\nn
\end{align}
This leads to Poincar\'e representations with both non-vanishing spin and mass.

%The constants $\cc_1$ and $\cc_2$ add a conformally invariant potential. In terms of cosmologies, they act as scalar field density, for which is known that the $\sl(2,\R)$ symmetries exist \cite{BenAchour:2019ufa, BenAchour:2019ywl} in the same time gauge as for the perfect fluid. Even if this might add an extra deformation of the trajectories, it does not alter the conclusions about the sign of the fluid densities, nor about the black hole singularity.

\medskip

The role of $\cc_{3}$ and $f(x)$ are pretty different from the one of $\cc_{1}$ and $\cc_{2}$. They are both generated by canonical transformations. The $\cc_{3}$-deformation is generated by the Poisson flow along $P_{2}$:
\be
\{P_{2},\cH^{(0)}\}=P_{2}
\,\,\Rightarrow\,\,
e^{\cc_{3}\{P_{2},\cdot\}}\cH^{(0)}
=
\cH^{(0)}+\cc_{3}P_{2}
\,.
\ee 
The $f(x)$-deformation is generated by the Poisson flow along $x=V_{2}/V_{1}$. Indeed, we compute the iterative Poisson brackets,
\be
\{x,\cH^{(0)}\}=-P_{1}
\,,\quad
\{x,P_{1}\}=-\f{V_{2}}{V_{1}^{2}}
\,,
\ee
with the next Poisson bracket vanishing. We can thus compute the general flow generated by $x$ on the original Hamiltonian:
\be
e^{\{\varphi(x),\cdot\}}\cH^{(0)}
=
\cH^{(0)}-\pp_{x}\varphi(x)P_{1}
+\f12x(\pp_{x}\varphi)^{2}\f1{V_{1}}
\,,
\ee
which gives the expected $f(x)$ correction terms with $f(x)=\pp_{x}\varphi(x)$.
Finally, the $\cc_{3}f(x)$ term in the full Hamiltonian is due to the non-vanishing Poisson of $P_{2}$ and $x$, which leads to a $1/V_{1}$ term:
\be
\left\{x,P_{2}\right\}=\f1{V_{1}}\,.
\ee
These canonical transformations affect the trajectories in a clear way and do not alter significantly the dynamics. At an intuitive level, the constant $\cc_3$ and the function $f$ couple linearly with the momenta $P_1$ and $P_{2}$,  they thus merely represent a shift of the canonical map between velocities $\dot V_{1,2}$ and momenta $P_{1,2}$, similar to Galilean boosts. We postpone the phenomenological analysis of such shifts to future investigation.

%%%%%%%%%%%%%%%%%%
\section{Conclusion \& Outlook}
%%%%%%%%%%%%%%%%%%%%%%%

The starting point of this paper was the study of the black hole mini-superspace, defined as the reduction of general relativity to spherically-symmetric metrics, here given by
\be
\de s^2 = 2N(r)\de r \de u + \f{\Vtwo(r)}{2\Vone(r)} \de u^2 + \ls^2 \Vone(r) \de \Omega^2
\,,\nn
\ee
where the metric components $N(r)$, $V_{1}(r)$ and $V_{2}(r)$ depend only on the radial coordinate $r$. Performing a canonical analysis for the evolution of those fields along $r$, the system admits a Hamiltonian formulation. The $g_{ru}$ component $N(r)$ is a Lagrange multiplier enforcing that the Hamiltonian vanishes and consists in a Hamiltonian constraint $\cH$, as usual in general relativity. The flow generated by this Hamiltonian constraint gives the reparametrization-invariant evolution along the proper radial coordinate $\tau$, defined as $\rd\tau=N(r)\rd r$.

Previous work \cite{Geiller:2020xze} identified a complete set of constants of motion, which were shown to generate a symmetry of the black hole mini-superspace, under the Poincar\'e  group $\ISO(2,1)$. This Poincar\'e group is not the metric isometry group and is not the group of asymptotic symmetry, but corresponds to non-trivial symmetry transformations in the field phase space, defined as Mobius transformations in the proper radial coordinate $\tau$ and their co-adjoint action \cite{Geiller:2021jmg}.
These conserved charges allow to fully integrate the motion of the system and play the role of the integration constants for the trajectories of the fields $V_{1}(\tau)$ and $V_{2}(\tau)$. The two Poincar\'e Casimirs vanish, so that the system correspond to a massless and spinless representation of the Poincar\'e group.

We embarked here on a systematic study of possible deformations of the dynamics of the system, compatible with the previously uncovered integrability structure. We identified a 5-parameter family of continuous deformations of the Hamiltonian constraint. Two parameters add terms proportional to the conjugate momenta of $V_{1}$ and $V_{2}$ and lead to shifts in their velocities. Two other parameters add potential terms in $1/V_{1}$ and $V_{2}/V_{1}^{2}$, which directly source the Poincar\'e Casimirs and give their non-zero values. The final parameter surprisingly leads to a deformation of the symmetry algebra, regularizing the non semi-simple Poincar\'e group to the semi-simple symmetry groups $\SO(3,1)$ or $\SO(2,2)$. We call this deformation parameter $\lambda$, for its similarity to the algebraic role of the cosmological constant in space-time isometries. We nevertheless underline that the $\lambda$-deformation of the Hamiltonian constraint has nothing to do with the cosmological constant volume term in the action of general relativity.

We focussed on the negative deformation parameter case, with $\lambda<0$, with its $\SO(2,2)$ symmetry group. We showed that the  deformed dynamics of the black hole mini-superspace can be represented as a superposition of two FRW cosmologies (for general relativity coupled to a homogeneous isotropic perfect fluid), and that the resulting modified black holes are non-singular space-times with a bouncing induced metric on the celestial sphere. These are very similar to singularity-avoidance scenarios for black holes and black-to-white hole transitions in quantum gravity, see e.g. \cite{Ashtekar:2005qt,Modesto:2005zm,Boehmer:2007ket,DeLorenzo:2015taa,Olmedo:2017lvt,Ashtekar:2018cay,Ashtekar:2018lag,BenAchour:2017ivq,BenAchour:2018khr,Bodendorfer:2019cyv,Bodendorfer:2019nvy,Alesci:2020zfi,Zhang:2021wex}. Our analysis suggests to revisit these various proposals in terms of symmetry and look for a universal symmetry-based argument for the resolution of the Schwarzschild black hole singularity.

Beside this main prospect, the present results open other doors.
First, now that the algebraic structure of the deformations of the black hole mini-superspace is settled, one could push further the study of this general model. At the classical level, one should perform a thorough analysis of the phenomenology of the resulting modified black hole geometries with $\lambda$-deformation, non-vanishing Casimirs and canonical shifts. Then, at the quantum level, one should perform a group quantization of the black hole phase space in terms of Poincar\'e and Lorentz representations and understand the coherence of semi-classical wave-packets of geometry can travel through the singularity-resolving bounce.

Second, the realization of the black hole dynamics as the superposition of two FRW cosmologies begs the question of superposing metrics and geometries in general relativity, a question that should then become essential in the perspective of writing a consistent quantum gravity theory. Superposition of states is a delicate issue in non-linear theories and it would be enlightening to understand if the present superposition mechanism can be generalized beyond spherically-symmetric or homogeneous metrics. This seems to echo the non-linear method of metric superposition used for cylindrically-symmetric space-time, e.g. to build black hole geometry with surrounding matter \cite{Chen:2023akf}, but more work is definitely needed to understand if there is a link between the present approach and those methods.

Finally, deforming the Hamiltonian constraint for the black hole mini-superspace is actually equivalent to modifying the Einstein equation for spherically-symmetric metrics. Here, the modifications are justified by preserving or regularizing the symmetry group. Focussing on symmetry and the algebra of conserved charges usually allow to keep a tight control over anomalies when quantizing the theory. One should definitely investigate if this fits with other symmetry-based works on modified black hole, e.g. using a modified Dirac algebra \cite{Alonso-Bardaji:2021yls}, and if our method can possibly be generalized beyond the reduction of general relativity to spherically-symmetric space-times.

\section*{Acknowledgement}

ER Livine would like to thank RIKEN's iTHEMS team (Wako, Japan) for its hospitality during the final stages of the research presented in this manuscript.

\bibliographystyle{bib-style}
\bibliography{biblioQG}
%%%%%%%%%%%%%%%%%%%%%

\end{document}